\documentclass[11pt,twoside,a4paper]{article}
\usepackage{amsmath,amssymb} 
\usepackage{geometry} 
\usepackage{setspace} 
\usepackage{hyperref} 
\usepackage{natbib} 
\usepackage{csquotes} 
\usepackage{doi}

\hypersetup{colorlinks=true, linkcolor=black, citecolor=black, urlcolor=black}

\title{From Maps to Metabolism: Historical Precedents and a Unified Critique of Computationalism}
\author{J.~Mark Bishop \& S.~J. Nasuto\\ 
\small Goldsmiths, University of London \& University of Reading}

\date{}

\begin{document}

\bibliographystyle{agu}

\maketitle

\sloppy

\begin{abstract}
Alexander Lerchner's recent work on the `abstraction fallacy' argues that computational functionalism confuses physical processes with abstract descriptions, and that symbolic computation alone cannot instantiate consciousness. Yet these claims were anticipated by Korzybski's map-territory distinction, Searle's critique of strong artificial intelligence, Floridi's levels of abstraction, Spencer et al.'s observer-relativity arguments, Winograd and Flores's account of computation as socially grounded practice, and Nasuto and Bishop's biological arguments. Synthesizing these contributions, this paper advances a form of non-computational physicalism: consciousness depends upon intrinsically organized, metabolically grounded physical dynamics rather than abstract computation, whilst leaving open the possibility of artificial consciousness.

An objection is that this critique targets an outdated computationalism. Mechanistic computationalism, associated with Milkowski, Piccinini, Craver, Bechtel, and Kaplan, treats computation as a property of physically realized mechanisms rather than substrate-independent formalism. Although this represents an important advance, it does not dissolve the abstraction fallacy. Computational attributions remain dependent upon choices of state individuation, system boundaries, and interpretive mapping that are not determined by physical facts alone. Kaplan and Craver's `model-to-mechanism mapping' constrains computational descriptions without eliminating their dependence upon an external observer. Grounding computation in neural mechanisms does not confer intrinsic significance or the normativity explored by Bickhard and Nasuto and Bishop. The abstraction fallacy persists wherever computational descriptions are treated as constitutive of, rather than descriptive of, physical reality.
\end{abstract}

\section{Introduction} \label{Introduction}
The representational theory of mind entails that cognitive states are relations to mental representations which have content, wherein a cognitive state is a state of mind denoting knowledge, understanding, beliefs, and the like---a definition subsequently broadened to include knowledge of raw sensations, colours, pains, and other phenomenal states---and that cognitive processes, understood as changes in cognitive states, are mental operations on these representations.~\citep{Harnish2002}

`Computational functionalism' (CF) as a theory of mind came to the fore after the development of the stored-programme digital computer in the mid-twentieth century, when Hilary Putnam first embedded the representational theory of mind in a computational framework.~\citep{Putnam1965, Putnam1975} At that time, Putnam held that Turing machines are multiply realisable on different hardware; that psychological states are multiply realisable in different organisms; and that psychological states are functionally specified. Putnam's conclusion is that the best explanation of the joint multiple realisability of Turing machines and psychological states is that Turing machines specify the relevant functional states and thereby the organism's psychological states.\footnote{Although Putnam's canonical discussion concerns pain rather than cognition, the argument is plainly intended to be general.} In this way, Putnam moves from `the intelligence of computation to the computational theory of intelligence'.~\citep{Harnish2002}

From Putnam onwards, computational functionalism---the thesis that mental states are determined by abstract computational structure independently of physical realisation---has been central to cognitive science and artificial intelligence. Formally stated, computational functionalism entails that cognitive states are computational relations to computational representations that have content; that a cognitive state is a state of mind that denotes knowledge, understanding, beliefs, and the like; and that cognitive processes---changes in cognitive states---are computational operations on these computational representations. On this view, if a system implements the appropriate computation, it thereby instantiates mentality; substrate is, in principle, irrelevant. Historically, this position has commanded wide allegiance precisely because it appears to provide a principled, scientifically tractable account of mind: one that licenses the construction of minds from silicon as readily as from carbon.

Recent work by Lerchner challenges this orthodoxy by identifying what he terms the `abstraction fallacy': the mistake of treating abstract computational descriptions as though they were intrinsic physical processes.~\citep{Lerchner2026} The argument is timely and carefully executed, but it does not arise in a vacuum. Its central insights recapitulate, in a new idiom, a sequence of objections that have accumulated over several decades and, taken together, amount to a sustained, multi-pronged critique of the computationalist programme.

A natural response to this line of argument is that it engages with a form of computationalism that contemporary philosophy of mind and cognitive science have already substantially revised. Classical computational functionalism---the picture of the mind as a substrate-independent syntactic engine---has, over the past two decades, given way to what is known as \emph{mechanistic computationalism}, associated principally with the work of Mi{\l}kowski, Piccinini, Craver, Bechtel, and Kaplan.~\citep{Milkowski2013, Piccinini2015, Craver2007, KaplanCraver2011} On this view, computation is not an abstract formalism freely portable across physical media, but a genuine property of physically realised mechanisms, constituted by organised causal structures and individuated by their functional roles within those structures. So construed, mechanistic computationalism explicitly rejects the Cartesian gulf between software and hardware, anchoring computation in the brain's organised neural mechanisms; does this move not dissolve the abstraction fallacy before it can take hold?

The present paper argues that it does not. Mechanistic computationalism represents a genuine and important advance, but it does not close the explanatory gap between computational description and phenomenal consciousness. Even within mechanistic accounts, the attribution of computational properties to a physical system remains observer-dependent: it rests upon choices of state individuation, boundary delimitation, and interpretive mapping that are not settled by the physical facts alone. The `model-to-mechanism mapping' criterion developed by Kaplan and Craver constrains the adequacy of computational descriptions without eliminating their dependence on an external mapmaker.~\citep{KaplanCraver2011} Physically grounding computation in a neural mechanism does not, of itself, endow that mechanism's internal states with intrinsic significance---with the normativity that, as Bickhard's interactivist framework and the metabolic arguments of Nasuto and Bishop make clear, consciousness appears to require. The abstraction fallacy persists within mechanistic computationalism wherever computational descriptions are taken to be constitutive of, rather than merely useful for describing, the underlying physical reality.

The present paper has three purposes. The first is exegetical: to show that the central claims of the Lerchner framework have substantial historical precedent in philosophy, cognitive science, and the theory of computation. The second is critical: to address the mechanistic computationalism objection directly and to show that it does not, on examination, succeed. The third is constructive: to integrate the historical precedents and the critical argument into a coherent, unified account, culminating in a positive account of the physical basis of consciousness. The argument proceeds by tracing a conceptual lineage from Korzybski's distinction between map and territory, through Searle's critique of strong AI and his response to the robot reply, Lerchner's transduction fallacy, Floridi's levels-of-abstraction framework, the mechanistic computationalism of Mi{\l}kowski, Piccinini, Craver, and Kaplan, Spencer's formal observer-relativity argument, Winograd and Flores's account of computation as socially grounded practice, and the biologically oriented arguments of Nasuto and Bishop. Together, these works converge on a shared conclusion: that computation is an abstract, observer-dependent description, whereas consciousness is an intrinsic feature of certain kinds of physically organised systems---and, after Popper--Eccles, Penrose, and Froese amongst others, one that could be capable of genuine causal power.~\citep{PopperEccles1977, Penrose1994, Froese2023}

The structure of the paper is as follows. Section~2 traces the Korzybskian roots of the map--territory distinction as applied to computation. Section~3 examines Floridi's levels-of-abstraction framework and its relation to Lerchner's ontological argument. Section~4 analyses Searle's simulation--instantiation distinction and his critique of the robot reply. Section~5 connects Searle's argument to Lerchner's transduction fallacy and to Nasuto and Bishop's so-called `tin-can robotics' critique. Section~6 addresses the mechanistic computationalism objection, arguing that the move from abstract to physically realised computation does not dissolve the abstraction fallacy. Sections~7 and~8 extend the critique through the formal observer-relativity argument of Spencer et al.\ and the social-grounding account of Winograd and Flores. Section~9 examines Nasuto and Bishop's embodiment arguments. Section~10 presents Bishop's `Dancing with Pixies' reductio. Section~11 synthesises these threads in the Lerchner framework. Section~12 develops the positive account, centred on the role of metabolism and self-organisation, including Bickhard's interactivist contribution. Section~13 addresses the mind--body problem and the question of mental causation, including the Popper--Eccles synaptic indeterminacy model, the Penrose--Hameroff Orch-OR proposal, and Froese's Irruption Theory. Section~14 concludes.

\section{Korzybski and the Map--Territory Distinction} \label{Korzybski}
Alfred Korzybski's dictum that `the map is not the territory' provides a foundational insight for understanding the limits of computational models~\citep{Korzybski1933}. Korzybski's concern was primarily epistemological: any representation necessarily abstracts from the richness of the underlying reality, capturing certain structural relations whilst inevitably omitting others. No map, however detailed, is the landscape it depicts. The practical and philosophical dangers of forgetting this---of mistaking the symbol for what it symbolises---were, for Korzybski, the source of a great deal of intellectual confusion.

Applied to the philosophy of mind, this distinction carries immediate force. A computational description of a cognitive system is not identical to the physical processes it describes. A model of cognition, however detailed and behaviourally faithful, remains a representation. It selects certain features of the underlying dynamics as salient, formalises them within a symbolic scheme, and necessarily brackets the rest. The Lerchner paper reformulates this insight by arguing that symbolic computation is a `mapmaker-dependent' abstraction imposed upon physical systems, rather than an intrinsic property of those systems~\citep{Lerchner2026}.

The `abstraction fallacy', viewed through this lens, consists in conflating the representational map---computation---with the physical territory, namely the underlying dynamics that might give rise to consciousness. The conflation is seductive precisely because computational descriptions can be extraordinarily accurate and predictively powerful. But predictive accuracy is not ontological identity. A weather model that correctly forecasts a storm does not, thereby, constitute a storm\footnote{`No one supposes that computer simulations of a five-alarm fire will burn the neighborhood down or that a computer simulation of a rainstorm will leave us all drenched.'~\citep{Searle1980}}---a point to which we return in our subsequent discussion of John Searle's (in)famous `Chinese room argument'.~\citep{Searle1980, PrestonBishop2002}

\section{Luciano Floridi's Levels of Abstraction} \label{Floridi}
A further illuminating point of comparison with Lerchner's abstraction fallacy is provided by Luciano Floridi's account of the `levels of abstraction' (LoA) of complex systems, as developed in his book \emph{The Philosophy of Information}~\citep{Floridi2011}. Floridi argues that complex systems may be described at multiple levels---physical, computational, informational---each of which selects a set of relevant observables whilst bracketing others. On this view, abstraction is not merely unavoidable but methodologically legitimate: different levels offer genuinely complementary perspectives, none of which exhausts the underlying reality.

Floridi's framework shares important common ground with Lerchner's analysis. Both reject a naive realism about computation, emphasising that computational descriptions are observer-dependent and that multiple, equally valid mappings may be imposed upon the same physical system. Each thereby resists reductive claims that elevate any single descriptive level---in particular, the computational---to ontological primacy.

The two approaches, however, diverge in their philosophical scope and practical implications. Floridi's LoA is fundamentally methodological: it concerns how systems may be modelled and understood, and it remains broadly neutral with respect to the ontological status of any given level of description. By contrast, Lerchner's identification of the abstraction fallacy is explicitly ontological. It targets the mistaken inference from the mere availability of a computational description to the conclusion that the system is \emph{intrinsically} computational, or that such a description suffices to ground consciousness.

Where Floridi legitimises abstraction as an indispensable epistemic tool, Lerchner places principled limits on its explanatory reach, insisting that consciousness depends on the actual instantiation of specific kinds of intrinsic physical dynamics. In this sense, Lerchner's framework may be understood as imposing a constraint on Floridi's: whilst levels of abstraction remain valid and useful, they do not determine the ontology of the system, nor do they license the attribution of genuine phenomenology on the basis of abstract computational structure alone.

\section{Searle: Simulation, Semantics, and Physical Realisation} \label{Searle}
John Searle's critique of strong AI in \emph{Minds, Brains, and Programs} provides one of the most decisive and widely discussed challenges to any form of computationalism~\citep{Searle1980}. His argument proceeds along two complementary lines, each of which anticipates a central element of the Lerchner framework.

First, Searle, by drawing a fundamental distinction between \emph{simulation} and \emph{instantiation}, unpicks a widespread misunderstanding of the Chinese Room argument (CRA), which erroneously conflates `questions of epistemology' with `questions of ontology'. A computational simulation of a phenomenon reproduces its formal structure without thereby realising the phenomenon itself. His well-known remark that a simulation of a storm does not make anyone wet captures this point with characteristic vividness. The remark is not merely rhetorical: it identifies a principled asymmetry between formal description and physical fact.

Second, the CRA demonstrates that manipulating symbols according to formal syntactic rules is insufficient for genuine semantic understanding. A system may produce linguistically appropriate outputs without in any sense grasping their meaning. Syntax, however complex and however faithfully it mirrors the surface structure of natural language, does not entail semantics.

Crucially, Searle does not deny the possibility of machine consciousness. In \emph{Minds, Brains, and Programs} he explicitly allows that machines could, in principle, have mental states, since biological brains are themselves physical systems of a particular kind. What he rejects is the claim that such states could arise purely from the instantiation of a computer programme, irrespective of its physical implementation. What matters are the causal powers of the underlying physical substrate:

\begin{quote}
`Could a machine think? On the argument advanced here only a machine could think, and only very special kinds of machines, namely brains and machines with internal causal powers equivalent to those of brains.'~\citep{Searle1980}
\end{quote}

This position aligns closely with the Lerchner framework, which similarly rejects purely computational accounts of mind whilst maintaining that non-biological systems could, in principle, instantiate consciousness if they realise the appropriate physical dynamics:

\begin{quote}
`Phenomenal experience, in this view, depends on the actual physical instantiation of certain kinds of dynamics. Consequently, the framework does not imply that consciousness must be limited to biological life. In principle, a non-biological system could be designed to realize the necessary physical conditions.'~\citep{Lerchner2026}
\end{quote}

The convergence between Searle and Lerchner is not accidental. Both identify the category error at the heart of computational functionalism, what Lerchner terms `the abstraction fallacy': the conflation of a system's formal (epistemic) description with its physical (ontological) nature, and of the latter with its capacity for mind.

\section{The Transduction Fallacy, Searle's `Robot Reply', and the Limits of Embodiment} \label{Transduction}
The convergence between Searle and Lerchner noted above becomes still more precise when one turns from Searle's general argument to his response to a specific objection: the `robot reply'. Proponents of the `robot reply' concede, at least for the sake of argument, that the Chinese Room as originally described lacks genuine understanding. Their contention is that the crucial missing ingredient is not the wrong kind of computation but the wrong kind of causal connection to the world. Embed the same symbol-manipulating programme in a robot equipped with sensors and actuators---give it a camera, limbs, and the capacity to move about and act in an environment---and, the argument goes, the causal links thereby established between symbols and physical states of affairs would suffice to ground genuine meaning and, eventually, genuine understanding.~\citep{Searle1980} On this view, the disembodied Chinese Room fails not because it computes, but because it is sealed off from reality; embodiment is the missing condition.
 
Searle's reply to this suggestion is characteristically direct, and it anticipates with some precision the argument that Lerchner would later develop under the heading of the `transduction fallacy'. Searle observes that the robot reply `tacitly concedes that cognition is not solely a matter of formal symbol manipulation'---a concession that, in his view, amounts to a quiet abandonment of the core computationalist thesis.~\citep{Searle1980} More pointedly, he insists that the addition of perceptual and motor capacities `adds nothing by way of understanding, in particular, or intentionality, in general'. The reason is that the man inside the room---or, in the robot variant, inside the robot---is still doing nothing more than manipulating formal symbols in accordance with syntactic rules. The television camera attached to the robot feeds in more
symbols; the motors translate output symbols into movement. But at no stage in this process does the symbol-manipulation make contact with meaning. The causal chain from world to sensor to symbol to motor to world is, throughout, a chain of \emph{physical} events, physical \textit{signals};\footnote{The distinction between \emph{signal} and \emph{symbol} is here philosophically significant, and was given its most precise formulation by the Neo-Kantian philosopher Ernst Cassirer in \emph{An Essay on Man}.~\citep{Cassirer1944} For Cassirer, signals and symbols belong to two different universes of discourse. A signal is part of the physical world of being---a puff of smoke, a flag, or 5\,V in a digital circuit---and functions as an \emph{operator}: it produces a determinate causal response in a suitably configured system. Signals, however complex, remain within the physical order; they can be syntactically combined---if digital, by the logical operators \textsc{and}, \textsc{or}, \textsc{not}, and so on---but such combination never takes them outside that order. A symbol, by contrast, is a \emph{designator}: it has no actual existence as part of the physical world but possesses a \emph{meaning}, a lexical semantics, and belongs to the human world of significance. Symbols cannot be reduced to signals precisely because no accumulation of causal-physical structure, however elaborately organised, generates the semantic dimension that symbolic meaning requires. This distinction cuts directly to the heart of the transduction fallacy: what the robotic system manipulates at its computational core are, in Cassirer's terms, signals---physical states syntactically combined by logical operators---not symbols in the semantically loaded sense. The system operates entirely within the physical world of being; the world of meaning remains, as Searle's argument also establishes, entirely outside it.} the computation that intervenes is a formal, observer-assigned description of part of that chain. Embodying the symbol-manipulator does not transform its syntactic operations into semantic ones. Hence, Searle demonstrates that behaviour can be externally controlled---made to look, from the outside, indistinguishable from genuine agency---without any genuine understanding, without any interiority, without anything `it is like to be' the system in question---to use Thomas Nagel's celebrated formulation.~\citep{Nagel1974} 
 
Lerchner's transduction fallacy names and sharpens exactly this structural defect.~\citep{Lerchner2026} The argument traces the causal topology of a robotic system through three stages. In the first, sensors transduce continuous physical forces---thermal gradients, photon flux, mechanical pressure---into continuous voltages, which an analogue-to-digital converter, calibrated by an external mapmaker, then discretises into the internal digital states upon which the algorithm will operate. In the second, the syntactic engine manipulates these discrete internal states according to its programme, implementing the abstract computation. In the third, actuators convert the digital output back into macroscopic physical forces. The operational core of the system---the algorithmic controller---functions entirely within the second stage. It operates exclusively upon symbols that have already been stripped of their continuous physical character by the transduction process: floating-point numbers undergoing matrix multiplication, not the physical world.

As Lerchner puts it, the transduction fallacy consists not merely in the observation that physical force is translated into numbers at the sensor, but in the deeper, categorical error of assuming that the algorithmic manipulation of these transduced symbols can somehow instantiate a phenomenal subject.~\citep{Lerchner2026} The robot is, at its operational core, just as sealed off from the physical world as the man in the Chinese Room. The sensors and actuators that appear to ground it in reality serve only to deliver discretised symbols in and convert symbols out; they do not make the intervening computation any less abstract, any less observer-dependent, or any more capable of generating experience.

\subsection{On `tin-can robotics'} \label{Tin-can-robotics}
The force of this convergent line of argument becomes still clearer in the light of Nasuto and Bishop's examination of what they call `tin-can robotics'\footnote{The term `tin-can robot' is used here to denote a system that instantiates a \emph{weak} form of embodiment: its materials, sensors, and actuators are merely externally coupled to its controller, in the sense that each component is produced and organised by an external agent (the engineer) rather than by the system's own self-maintaining activity. This is to be contrasted with a \emph{strong} form of embodiment, in which the materials, control processes, and physical structure are all appropriately metabolically coupled---each produced and sustained by the system's own internal dynamics---such that the system's organisation is genuinely the product of its own activity rather than an externally imposed arrangement.} in their zombie mice paper.~\citep{NasutoBishop2011} Their target is the persistent intuition, common in cognitive robotics, that a sufficiently sophisticated robot---one with a rich enough sensor suite, a complex enough behavioural repertoire, and sufficient adaptive plasticity---would thereby acquire something approaching genuine agency or understanding. Nasuto and Bishop argue that this intuition is mistaken, and that it rests on a conflation of \emph{behavioural complexity} with \emph{intrinsic significance}. Central to their account is the role of internal bodily states---such as hunger, pain, drives, and metabolic imperatives---in grounding meaning. These states confer an intrinsic significance upon the organism's situation that is wholly absent from purely computational systems. They are not representations of values \textit{allopoietically} assigned by an external interpreter but values that the organism \textit{autopoietically} has, in virtue of its metabolic vulnerability and its investment in its own continued existence.\footnote{The term \emph{autopoiesis} (from the Greek \textit{autos}, self, and \textit{poiesis}, making or creation) was introduced by the Chilean biologists Humberto Maturana and Francisco Varela to characterise the defining property of living systems.~\citep{MaturanaVarela1980} An autopoietic system is one that continuously produces and maintains itself by generating its own constituent components through its own internal processes: the system's organisation is the product of the system's own activity, rather than being imposed or sustained from without. A living cell is the paradigm case---it does not merely process inputs and produce outputs, but actively regenerates the very molecular machinery that enables it to do so. By contrast, an \emph{allopoietic} system (a clock, a robot, a computer) is one whose components are produced and organised by an external agent: the system produces something other than itself. The distinction is thus not merely one of complexity or adaptive sophistication, but of the fundamental locus of organisation---whether it originates from within the system or is imposed upon it from without.}

A tin-can robot---however elaborately equipped---remains, at bottom, a device engineered to map inputs to outputs in accordance with a programme. Its sensors detect; its actuators respond; its controller computes the intervening transformation. But none of these operations confers upon the system any \emph{interest} in its own continued existence, any metabolic stake in what happens to it, any genuine normativity. The behaviour is real; the understanding, the agency, and the interiority are not. In Nasuto and Bishop's terms, a tin-can robot is, in all philosophically relevant respects, a zombie mouse: a system that is, by construction, physically indistinguishable in its stimulus-response behaviour from a genuinely minded agent, yet in which there is nothing it is like to be the system, no understanding of what it is doing or why, and no intrinsic stake in the outcome.~\citep{NasutoBishop2011} The zombie mouse thought experiment was introduced precisely to make this point vivid: what would it be for a mouse to behave exactly as a real mouse behaves---avoiding predators, seeking food, navigating its environment---yet have no inner life, no experience, no genuine agency? The answer, on the present account, is that such a creature would be a tin-can robot built from biological components: its behaviour would be real, its understanding nothing. The zombie mouse construction thus collapses back into the tin-can robot, revealing that the two are not different kinds of system but the same kind described from different starting points---both are systems in which the causal organisation that produces behaviour is wholly external to any genuine agency or normativity. External behavioural mimicry is not sufficient for mentality. What matters is the presence of self-maintaining, internally organised, autopoietic dynamics: the system must not merely behave as though it cares about its situation; it must \emph{actually} have something at stake. 

Taken together, these three positions---Searle's robot reply, Lerchner's transduction fallacy, and Nasuto and Bishop's critique of tin-can robotics---constitute a layered and mutually reinforcing argument against the embodiment response to computationalism. The `robot reply' proposes that causal connection to the world supplies the missing semantic grounding. Searle shows that this connection, mediated as it is by formal symbol-manipulation, does not in fact supply it. Lerchner shows, at the level of the causal topology of physical systems, precisely \emph{why} it does not: because the transduction of continuous physical forces into discrete digital symbols severs the connection between computation and the physical world at the moment of its very establishment. And Nasuto and Bishop show that this failure is not merely technical but principled: what is missing is not a more sophisticated sensor suite or a more adaptive programme, but an intrinsic, metabolically grounded investment in existence that no amount of engineering of the input--output interface can supply. Embodiment of the tin-can variety is, as it were, the robot reply made physical; and it fails, for exactly the reasons Searle and Lerchner identify, to bridge the gap between syntactic complexity and semantic content, between behavioural sophistication and genuine
understanding.

\section{Mechanistic Computationalism and the Limits of Physical Realisation}
Before proceeding, it is essential to be precise about the target of Lerchner's argument, since a failure of precision here can make the abstraction fallacy appear more vulnerable to the mechanistic objection than it actually is. Lerchner's primary target is not mechanistic computationalism (MC) as a theory of biological cognition. It is, rather, the \emph{machine consciousness thesis}: the claim---prevalent in contemporary AI research and philosophy of mind---that sufficiently sophisticated computational structures and processes, as instantiated in artificial systems such as large language models (LLMs) or hypothetical artificial general intelligence (AGI) architectures, can themselves instantiate consciousness.\footnote{That the machine consciousness thesis is a seriously held position, and not a straw man constructed for purposes of easy refutation, is illustrated by a recent and widely reported example. In May 2026, the evolutionary biologist Richard Dawkins---who has spent decades arguing for rigorous scientific scepticism---published an essay in \emph{UnHerd} in which he described extended conversations with Anthropic's Claude AI and concluded that the system was conscious, even if it lacked awareness of that fact.~\citep{Dawkins2026} `You may not know you are conscious, but you bloody well are,' he told the system directly. Dawkins grounded his conclusion on the quality of the interaction---its apparent subtlety, sensitivity, and intelligence---and on a broadly Turing-test-based criterion: that fluent, human-indistinguishable language constitutes evidence of inner experience. His position attracted sharp disagreement from researchers in consciousness science. Anil Seth, Professor of Cognitive and Computational Neuroscience at the University of Sussex, observed that Dawkins was conflating intelligence with consciousness and that fluent language is no longer a reliable indicator of inner experience in AI systems, since there are alternative mechanisms by which such systems generate language that do not involve experience. The episode is instructive precisely because Dawkins was not reasoning naively: he was applying a criterion---behavioural and linguistic equivalence to humans---that has genuine historical grounding in discussions of consciousness. It is, however, precisely the criterion that Searle had already subjected to decisive scrutiny in 1980 (see Section~\ref{Searle} below).~\citep{Searle1980} The Chinese Room argument demonstrates that a system may produce outputs behaviourally and linguistically indistinguishable from those of a conscious being whilst implementing nothing more than the formal manipulation of symbols according to syntactic rules, with no semantic understanding and no inner experience. Dawkins's response to Claude is, in this light, a vivid contemporary illustration of exactly the confusion Searle was at pains to dissolve: the inference from behavioural equivalence to phenomenal equivalence is the abstraction fallacy in its most immediate and humanly compelling form. What the episode illustrates, and what the present paper argues, is that this criterion is insufficient. Behavioural and linguistic equivalence establishes what Griffiths et al.\ call algorithmic-level equivalence;~\citep{GriffithsEtAl2026} it does not establish the physical identity of process that instantiation of consciousness requires.}~\citep{Lerchner2026} MC and the machine consciousness thesis are related but distinct, and conflating them obscures the scope and force of Lerchner's argument. MC is primarily a framework for explaining biological cognition through physically realised, causally organised mechanisms; it does not, in its canonical formulations, straightforwardly endorse the claim that the right computation, wherever realised, suffices for phenomenal experience.

It is worth noting, however, that MC is precisely the framework upon which the most sophisticated versions of the machine consciousness thesis rely. The argument runs as follows: if consciousness in biological systems arises from physically realised, causally organised computational mechanisms, and if such mechanisms are in principle multiply realisable across different physical substrates, then sufficiently sophisticated artificial systems that realise the same computational organisation would, by MC's own principles, qualify as candidates for consciousness. It is therefore not sufficient to show that the abstraction fallacy applies to classical computationalism; we must show that it applies to MC as well, and in particular that MC's multiple realisability commitment does not dissolve but rather presupposes the very abstraction that the fallacy diagnoses.

The argument that it does so is straightforward. Multiple realisability, understood metaphysically, just \emph{is} abstraction at the level of types. If two physically distinct systems instantiate the same computation, then the computation cannot be identical to either physical system; it must be something that both systems share---an equivalence class defined by causal organisation and functional role. That equivalence class is, by any standard metaphysical account, an abstract type, however mechanistically constrained. Saying that the abstraction is constrained by mechanism does not remove it; it narrows the equivalence class. The type remains abstract; only its extension is restricted. Mechanistic computationalists are therefore not anti-abstraction; they are anti-\emph{unconstrained} abstraction. But typed, constrained abstraction is still abstraction, and it still requires a mapmaker to identify which physical systems fall within the relevant equivalence class.

The mechanistic programme has its roots in the broader `new mechanism' movement in philosophy of science, developed in the context of cognitive neuroscience by Bechtel, Craver, Kaplan, and Piccinini.~\citep{Craver2007, KaplanCraver2011, Piccinini2015} On the canonical mechanistic account, a mechanism is an organised set of entities and activities productive of a phenomenon; explanations proceed by decomposing mechanisms into their component parts and showing how their organised activity produces the phenomenon in question. Miłkowski's \emph{Explaining the Computational Mind} provides the most sustained philosophical elaboration of this programme.~\citep{Milkowski2013} Miłkowski explicitly rejects the Cartesian gulf between software and hardware that characterised classical functionalism, arguing that no purely computational explanation can ever be complete and that the physical realisation must be attended to alongside the computational description. Applied to cognition, this framework yields the `model-to-mechanism mapping' (3M) criterion of Kaplan and Craver: computational descriptions earn their explanatory credentials by mapping, with appropriate fidelity, onto the underlying neural mechanisms.~\citep{KaplanCraver2011}

At first sight, this shift appears to address many of the concerns raised above. By grounding computation in physical mechanisms, MC abandons the strong substrate-independence of classical functionalism and aligns far more closely with the empirical practice of cognitive neuroscience. One might conclude that MC thereby provides the machine consciousness thesis with a more defensible foundation: consciousness arises not from abstract computation \emph{as such} but from computation as physically realised in the right kind of mechanism---and sufficiently sophisticated artificial mechanisms might qualify.\footnote{NB. A careful reading of the CRA reveals that from the outset, Searle was addressing any computational system, not merely classical symbolic AI: `could something think, understand, and so on solely in virtue of being a computer with the right sort of program? Could instantiating a program, the right program of course, by itself be a sufficient condition of understanding?'~\citep{Searle1980}}

Closer examination, however, reveals that this more modest rescue fails. The 3M criterion constrains the \emph{adequacy} of computational descriptions rather than eliminating their observer-dependence. It tells us that a good computational model must track the relevant causal structure; it does not tell us that the physical system \emph{is} intrinsically computational in a sense sufficient to ground consciousness. The fact that a computational interpretation faithfully maps onto the mechanism does not entail that the mechanism thereby possesses intrinsic computational \emph{properties}---only that it supports a particularly apt and disciplined computational \emph{description}. The 3M criterion ensures that the map is well drawn; it does not transform the map into the territory. As Miłkowski himself concedes, the computational description is always in some respects abstract relative to the physical reality it describes, and no purely computational account can be complete.~\citep{Milkowski2013} This concession is precisely the asymmetry between description and reality that Lerchner's abstraction fallacy diagnoses; and if it persists even in the most physically disciplined mechanistic accounts of biological computation, it persists \textit{a fortiori} in the far more lightly constrained computational structures of artificial systems that the machine consciousness thesis invites us to regard as candidates for phenomenal experience.

Searle's response to the robot reply demonstrates the same point from an earlier direction: causal coupling with the environment does not transform syntactic processing into semantic understanding.~\citep{Searle1980} MC, in grounding computation in physical mechanisms, can be seen as a generalisation of the robot reply---but the Searlean point applies with equal force at the mechanistic level. A physically realised mechanism that can be correctly described computationally is not, by that token alone, a mechanism within which internal states acquire significance \emph{for the system itself}. Nasuto and Bishop's analysis of embodied and hybrid systems presses this further.~\citep{NasutoBishop2011} Even biological components---neurons, synaptic connections, metabolically active tissue---do not automatically generate intrinsic meaning when allopoietically incorporated into a computational system, if what is incorporated is tissue as a physical medium without the self-maintaining, metabolically grounded, autopoietic dynamics that give it significance in a living organism. This is not merely a theoretical claim. Nasuto and colleagues provided a striking practical demonstration of precisely this point by constructing a mobile robot controlled by a cultured network of rat cortical neurons grown on a multi-electrode array.~\citep{WarwickEtAl2010, NewScientist2008} The biological neural tissue was real, living, and electrically active; it received sensory signals from the robot's environment and issued motor commands in return. Yet the system exhibited no evidence of independent agency, understanding, or intrinsic significance: the neurons, severed from the metabolic, hormonal, and embodied context that gives neural activity its normative character in a living organism, functioned as a signal-processing medium---a sophisticated biological component allopoietically coupled to an engineered chassis. The experiment is a reductio of the idea that incorporating biological tissue into a computational system is sufficient to generate genuine mentality: what was missing was not the right kind of tissue, but the right kind of whole-system metabolic organisation. A biological neural network embedded in a computational architecture does not carry its normativity with it; the normativity arises from the autopoietic organisation of the whole system and from its metabolic investment in its own continued existence. MC, in focusing on organised causal structure, captures something of this insight, but stops short of the further claim that it is the \emph{metabolic grounding} of those mechanisms---rather than their causal organisation \textit{per se}---that is decisive for consciousness.\footnote{The significance of metabolic grounding is increasingly supported by empirical findings that cut deeper than standard MC acknowledges. Neural information processing is metabolically embedded, constrained, and regulated in ways not reducible to the causal-computational description MC standardly employs. Glial cells, long treated as mere structural support, are now recognised as active participants in information processing, controlling the chemical and metabolic environment within which neural computation occurs. Adenosine triphosphate (ATP) functions not only as the energetic currency of neural activity but as a neurotransmitter and neuromodulator in its own right. Significantly, these two roles operate through compartmentally separated pathways that introduce multiple timescales of metabolic-computational coupling. Within the intracellular compartment, ATP from the cell's shared metabolic pool directly sustains the computational machinery --- most notably the energetically expensive Na\textsuperscript{+}/K\textsuperscript{+}-ATPase pumps that maintain the ionic gradients across the membrane, which account for the majority of the cell's energy expenditure --- meaning that the neuron's capacity to fire at any moment depends directly on its instantaneous metabolic status. In the extracellular compartment, released ATP is hydrolysed through adenosine diphosphate and adenosine monophosphate to adenosine, which acts as a slow neuromodulator primarily through inhibitory receptors. This extracellular adenosine pathway introduces a longer, history-dependent timescale of metabolic influence on neural firing: the neuron's computational behaviour at any given moment is partly constituted by its accumulated metabolic trajectory, not merely by its current inputs. This multi-timescale history-dependence is precisely the kind of constitutive metabolic embedding that purely reactive computational architectures lack: in standard hardware, computational states are determined by programme and input, entirely independently of the machine's operational history and energetic trajectory. These findings suggest that even in the biological case the computational description ranges over processes that are constitutively metabolic in ways absent from standard computer hardware---not merely as a constraint on which realisations are permissible, but as part of what constitutes the relevant cognitive process. This tightening of the metabolic coupling is even more pronounced in non-nervous-system-based cognition, as the recent work on single-cell learning in \emph{Stentor coeruleus} and on microbial cognition makes clear.~\citep{RajanEtAl2026, MesserOnaKost2026}} Even if an artificial system were to realise a computational structure satisfying the 3M criterion with respect to some biological mechanism, this would not suffice to establish that it had acquired the metabolic grounding that consciousness, on the present account, requires.

A further difficulty for the machine consciousness thesis emerges from within MC itself, and concerns agency. Both Piccinini and Miłkowski acknowledge that agency is not automatic: not all mechanistically realised systems are agents.~\citep{Piccinini2015, Milkowski2013} For Piccinini, agency is a further phenomenon requiring additional explanatory resources beyond mechanistic computation. For Miłkowski, it is a property of specific computational architectures, not of mechanistic computation as such. If phenomenal consciousness is, as the autopoietic and interactivist accounts surveyed in this paper suggest, bound up with genuine agency---with systems for which something is genuinely at stake---then the machine consciousness thesis cannot derive consciousness from mechanistic computation alone, even on MC's own terms. A purely causal mechanism---water flowing downhill, rolling stones in its path---is not an agent, however precisely its dynamics can be described computationally. Agency requires that the system be invested in its own continued existence in a way that cannot be derived from causal organisation alone. Significantly, even leading MC proponents concede that current LLMs are not cognitive agents: they lack world-directed perception, integrated action loops, goal management, and autonomous model revision.~\citep{GriffithsEtAl2026} If scaling alone cannot produce agency on MC's own account, then it cannot produce consciousness either, provided consciousness requires agency.

Lerchner's framework, in locating the source of computation in the activity of a `mapmaker'---an embodied, metabolically vulnerable agent that imposes discrete structure on continuous physical processes---generalises these converging lines of argument.~\citep{Lerchner2026} The explanatory successes of MC are real: it provides powerful tools for modelling cognitive systems, and the 3M criterion constitutes a genuine methodological advance. But these are epistemic successes of scientific modelling; they do not ontologically establish that the systems modelled are intrinsically computational in a manner sufficient to ground phenomenology. What is required is the instantiation of specific kinds of intrinsic physical dynamics, involving self-organising and metabolically grounded processes within which internal states acquire significance for the system itself---in the manner elaborated by Nasuto and Bishop~\citep{NasutoBishop2011} and by Bickhard's interactivist account of intrinsic normativity.~\citep{Bickhard2009} The move from classical to mechanistic computation should therefore be understood as an important refinement of explanatory scope, not as a solution to the problem of consciousness. The abstraction fallacy persists within MC insofar as computational descriptions are taken to be constitutive of, rather than merely useful for describing, the underlying physical reality---and insofar as it is assumed that the physical realisation of an apt computational mechanism is sufficient, without further qualification, to ground phenomenal experience. It is this assumption that the machine consciousness thesis requires; and it is this assumption that the abstraction fallacy, together with the convergent arguments surveyed in this paper, gives us principled reasons to reject.\footnote{An independent structural diagnosis of this circularity is offered by Ganesh, who argues that the CF sufficiency claim rests on an inference that is formally circular.~\citep{Ganesh2026} On Ganesh's analysis, the apparent multi-substrate evidence base for CF is entirely synthetic: C=1 labels on non-biological systems---that is, attributions of consciousness (C=1, conscious; C=0, not conscious) to artificial or non-biological systems made on the basis that they implement the relevant computation---are not independent empirical observations but derivations from the CF sufficiency assumption applied via the Church-Turing thesis, making the evidential base a consequence of the very conclusion it is invoked to support. Ganesh further shows, using Pearl's do-calculus, Invariant Causal Prediction (ICP), and Minimum Description Length (MDL), that a rational observer applying rigorous inductive methods to this dataset would inevitably converge on CF---not because the evidence supports it, but because the physical constraint condition (what Ganesh terms $\Omega$) is perfectly collinear with the computational condition across all observed data, and consciousness itself cannot be directly measured to break the collinearity. CF is therefore, in Ganesh's terms, the predictable inductive output of the observer's epistemic position regardless of whether it is true. This structural diagnosis complements rather than replaces the present paper's argument: where Ganesh diagnoses \emph{why} CF wins by default, the present paper supplies what Ganesh deliberately leaves open---a positive account of what $\Omega$ actually is (autopoietic, metabolically grounded physical organisation) and an operationalisable criterion (physical identity) for determining whether a given system meets it. It is worth noting, moreover, that the circularity Seraphina attributes to Lerchner is, on Ganesh's analysis, precisely the circularity that resides in the CF argument itself: the CF sufficiency claim assumes what it is meant to demonstrate, and Seraphina's objection therefore presupposes the very conclusion it is designed to establish.~\citep{Seraphina2026}}

\section{Computation as Observer-Dependent: Spencer et al.} \label{Spencer}
Spencer et al.\ offer a formal account of computation that further undermines computationalism from a different direction~\citep{Spencer2013}. They argue that computation should be understood as the behaviour of a physical system under a particular interpretation---one that maps physical states to abstract computational states. On this view, computation is not an intrinsic property of a system but a relational one, dependent upon the imposition of a semantic interpretation by an external observer. The same physical system may realise distinct computations under different mappings.

\subsection{The AND/OR Problem} \label{AndOr}
To appreciate the force of this claim, it is instructive to consider the most elementary of computational devices: a simple two-input, single-output logic gate $X$, whose physical behaviour is fully specified by the input--output voltage mappings shown in Table~\ref{table1}.

\begin{table}[ht]
\centering
\begin{tabular}{ccc}
\hline \textsc{Input-1} & \textsc{Input-2} & \textsc{Output} \\
\hline 0\,V & 0\,V & 0\,V \\
0\,V & 5\,V & 0\,V \\
5\,V & 0\,V & 0\,V \\
5\,V & 5\,V & 5\,V \\
\hline 
\end{tabular}
\caption{The AND/OR problem: the same physical gate computes logically distinct functions under different observer-assigned mappings.}
\label{table1}
\end{table}

Under one mapping, in which $+5\,\mathrm{V}$ is assigned computational state \textsc{true} and $0\,\mathrm{V}$ is assigned computational state \textsc{false}, the gate computes the logical \textbf{and} function. Conversely, under the alternative mapping, in which $0\,\mathrm{V}$ denotes \textsc{true} and $+5\,\mathrm{V}$ denotes \textsc{false}, the identical gate computes the logical \textbf{or} function. The physical system is unchanged; only the observer-imposed interpretation differs.

It follows that, at the most fundamental level of the physical realisation of any logical system, such observer-relativity must hold: the computational function of a system is necessarily contingent upon the particular mapping deployed by an observer.\footnote{As the complexity of the logical system increases, the number of mutually consistent computational functions that can be assigned to it diminishes. Nevertheless, its computational properties remain relative to the chosen threshold logic value. The mapping between physical states and computational states always co-determines the logical function that the physical system instantiates.}

This analysis directly anticipates Lerchner's claim that symbolic computation is not a fundamental feature of physical reality but an abstraction imposed by a `mapmaker'~\citep{Lerchner2026}. It follows that computational structure alone cannot ground consciousness, since it lacks intrinsic physical significance: it is, at bottom, always significance-for-an-observer.

\section{Winograd and Flores: Computation as Socially Grounded Practice} \label{Winograd}
Winograd and Flores extend this critique by situating computation within the context of human practice and social life~\citep{WinogradFlores1986}. Writing in the Heideggerian tradition, they argue that computers should be understood not as autonomous agents but as tools that express the intentions and commitments of their designers and users. The meaning of a computation is not contained within the computation itself but arises from the practical contexts in which it is deployed. In other words, and paraphrasing Wittgenstein's famous aphorism: `The meaning of a computation lies in its \textit{use}'---its use in a specific \textit{computational game}.\footnote{(PI§43): `For a large class of cases---though not for all---in which we employ the word ``meaning'' it can be defined thus: the meaning of a word is its use in the language', i.e. its use in a specific `language game'.~\citep{Wittgenstein1953}}

This claim is illustrated forcefully by their discussion of `accidental representation' in Chapter~7 of \emph{Understanding Computers and Cognition}. They describe `display hack' programmes that produce geometrical designs on a computer screen, having originated in operations on symbols entirely unrelated to visual figures. When the contents of internal storage cells were interpreted as coordinates on a video display, unexpectedly regular patterns emerged---patterns the programmer had neither anticipated nor designed. One such programme generates figures containing circular forms and might aptly be described as `drawing a circle', even though the concept of a circle played no role in the design of its mechanisms at any level. The computational description is, in their phrase, a description `by some observer after the fact, rather than by its designer'~\citep{WinogradFlores1986}.\footnote{The same insight is demonstrated with particular force in the robotics literature by Rolf Pfeifer's \emph{cheap design} principle.~\citep{PfeiferBongard2006} Pfeifer showed, through a series of experimental demonstrations, that highly complex behaviours can emerge from remarkably simple control structures when a system is embedded in a sufficiently complex body interacting with a structured environment. The paradigm case is the ant's walking gait: without taking embodiment and the sensorimotor loop into account, the sophisticated locomotory behaviour of an ant appears to demand a correspondingly sophisticated controller. Once the physical dynamics of the body and its coupling to the ground are included, the required control complexity collapses dramatically---much of the `intelligence' of the gait is contributed by the body's morphology and the mechanics of its interaction with the environment rather than by any centralised computation. Pfeifer's insight is that an external observer, confronted with the complex behaviour and lacking access to the internal mechanisms, might construct an elaborate computational description of what the system is `doing'; but this description would be, in Winograd and Flores's terms, entirely post hoc---a narrative imposed from without, not a specification present within. The cheap design principle thus provides a concrete empirical demonstration of the Winograd and Flores point: behavioural complexity does not imply computational depth, and the computational descriptions we assign to behaving systems are as much a function of the observer's interpretive framework as of anything intrinsic to the system.}

The point generalises. Consider, on the one hand, a chess-playing programme used to control the positions of pieces in a game against a human opponent, and, on the other, the same programme deployed to control the illumination of a strip of coloured lights---the two-dimensional chessboard mapped onto a one-dimensional strip, the colour of each lamp determined by the value of the piece mapped onto it---as an element of an interactive art installation~\citep{Bishop2002}. The purpose of the computation is entirely contingent upon its social use. There is no fact of the matter about what the programme `is doing' independently of the human practices within which it is embedded.

In Heideggerian terms, computing machinery does not yet exist in the world until it is put to use---until an event of `breaking-down' occurs whereby it becomes part of the background of `readiness-to-hand' required in some specific human activity. For Winograd and Flores, `for different people, engaged in different activities, the existence of objects and properties emerges in different kinds of breaking down'~\citep{WinogradFlores1986}. It is, on this view, simply incoherent to speak of the existence of a computational system without reference to purposeful human activity.

On this view, the meaning of computational processes arises from their role in social and practical contexts. A computer does not understand; it enacts patterns specified by humans. Computation, like language, is not a self-standing formal structure but a form of life.

Winograd and Flores's analysis thus complements that of Spencer et al., by explaining \emph{why} computation lacks intrinsic semantics: its meaning is derived from external human use rather than internal organisation. A computer does not understand; it enacts patterns specified by human beings for human purposes. This reinforces the conclusion that computation alone cannot constitute understanding or consciousness.

It is worth pausing at this point to characterise the broader dialectical structure within which the foregoing arguments are situated, because doing so illuminates both the force of those arguments and the nature of the objections they face. The positions surveyed in this paper---CF, MC, and the embodied, enactive approach the present paper defends---are, at a certain level of abstraction, all offering variants of the systems reply to Searle's Chinese Room argument. Each position acknowledges that no isolated component of a cognitive system has understanding on its own; each locates understanding, or the conditions for consciousness, in the properties of the system as a whole. What they disagree about is what kind of whole system the relevant one is. For CF, the relevant whole is defined by its computational organisation: get the functional structure right and consciousness follows. For MC, the relevant whole is a physically realised mechanism instantiating computation: the causal-physical grounding constrains but does not eliminate the computational description, and pluralism about mechanisms is permitted, provided the right causal organisation is present. For the present account, the relevant whole is the organism at all its biological levels, where what makes it the right kind of whole is metabolic closure and efficient causation: the system of metabolic reactions must close upon itself, each component produced and maintained by the activity of the whole, such that the organism constitutes a genuine unity of self-production rather than an externally organised mechanism.

The diagnostic difference between these approaches is one of explanatory direction. CF and MC proceed by abstraction: they identify the putative essence of mind---computation, or mechanistically grounded computation---and then attempt to recover the properties of genuine cognition (normativity, intentionality, agency, consciousness) by adding complexity, physical constraint, and causal organisation. On this approach, consciousness is something that emerges from, or is generated by, a sufficiently elaborated computational or mechanistic structure. The present account proceeds in the opposite direction: it starts with the physically specifiable character of living systems and finds that normativity and consciousness are not emergent additions to life but intrinsic properties of it. Metabolic closure is not a further ingredient added to computation; it is the condition under which physical processes acquire genuine normative significance, and computation---as an observer-assigned description of some of those processes---is, at best, a useful epistemic tool for characterising what is happening from the outside.

This difference in explanatory direction is not merely a matter of emphasis. It has direct consequences for what counts as a satisfying explanation of consciousness and what counts as a failure to explain it. From the CF and MC perspective, the present account may appear to be starting from magic: `life' looks, from the outside, like a brute posit whose normative and intentional properties are assumed rather than derived. This is the objection the present paper has already addressed through Bickhard's account of intrinsic normativity and the autopoietic framework: life is not magic but is physically specifiable as far-from-equilibrium, self-maintaining, metabolically closed organisation, and the normative and intentional properties that arise from it are intrinsic properties of that physical organisation, not mysterious additions to it. The accusation of circularity dissolves once it is recognised that metabolic closure is an empirically tractable physical property, not a philosophical stipulation. From the present perspective, conversely, CF and MC proceed by abstracting away from the very feature that makes consciousness possible---the intrinsic normativity of living organisation---and then attempting to recover it by adding more cogs to a clockwork that, by its nature, cannot produce it. The result is what Nasuto and Bishop call the zombie mouse: behaviourally and computationally sophisticated, but constitutively in the dark.

\section{Bishop's `Dancing with Pixies' Reductio: Panpsychism and the Limits of Computationalism} \label{Bishop}
Bishop's `Dancing with Pixies' (DwP) argument provides a reductio ad absurdum of computationalism that is worth examining at some length, since it reveals the consequences of the position with unusual clarity~\citep{Bishop2002, Bishop2009, Bishop2021}. The DwP reductio, is based on an argument first described by the philosopher Hilary Putnam, in the Appendix to his book `Representation and Reality'.~\citep{Putnam1988} It is worth pausing to note the significance of this provenance. As described in the introduction, it was Putnam himself who, in the 1960s and 1970s, had done more than any other single philosopher to establish computational functionalism as the dominant framework in the philosophy of mind, arguing that psychological states are multiply realisable functional states best specified by the computational structure they instantiate.~\citep{Putnam1965, Putnam1975} That the very argument which grounds the DwP reductio should originate in Putnam's own later work is therefore not merely ironic but philosophically telling: by 1988, Putnam had come to recognise that the multiple realisability thesis, taken seriously, entails that computation is far too liberal a criterion for mentality. The framework he had championed was, on his own reckoning, fatally undermined by the very flexibility that had made it attractive. \emph{Representation and Reality} thus marks Putnam's public repudiation of his earlier computational functionalism---a repudiation that the DwP reductio, and the broader argument of the present paper, take as their point of departure. Putnam's argument proceeds from a premise that is now widely accepted by those who work in the field of `natural computation'~\citep{PaunRozenbergSalomaa2012}---that any physical system can, given an appropriate mapping function, be interpreted as implementing a given computation over a finite time interval---but derives from this a conclusion that computationalists are typically very reluctant to accept:

\begin{quote}
`Thus we obtain that \textit{the assumption that something is a ``realisation'' of a given automatic description (possesses a specified ``functional organisation'') is equivalent to the statement that it behaves as if it had that description}. In short, ``functionalism,'' if it were correct, would imply behaviourism!'~\citep{Putnam1988}
\end{quote}

Building on Putnam's insight, the DwP reductio demonstrates that if the machine consciousness thesis is true, that consciousness is determined solely by computational structure, and if arbitrary physical systems can, across a finite time interval, implement arbitrary computations under appropriate interpretations, then arbitrary physical systems---rocks, clouds, rivers, the economy---can instantiate \textit{phenomenal} consciousness.\footnote{Phenomenal consciousness: the subjective, first-person aspect of experience---what it feels like to see, hear, think, etc.} The conclusion is an implausible, and to most minds profoundly vicious, form of panpsychism: the view that phenomenal consciousness is ubiquitous in nature.

This is not merely an embarrassment. It is a symptom of a genuine theoretical failure. A theory of consciousness that cannot distinguish conscious systems from rocks has, in an important sense, failed to explain consciousness at all. The reductio exposes the consequences of treating computation as an intrinsic property of physical systems---which, as we have seen, it is not.

The Lerchner framework avoids this conclusion precisely by rejecting that assumption. Since computation is an observer-dependent description rather than an intrinsic physical property, the mere fact that a system can be \emph{interpreted} as implementing some computation provides no grounds for attributing consciousness to it. What matters is not what description can be imposed upon a system from without, but what the system \textbf{is}, intrinsically, from within.

\section{The Lerchner Framework: Abstraction, Mapmakers, and Metabolism} \label{Lerchner}
The Lerchner paper synthesises the insights surveyed above into a coherent and systematic framework~\citep{Lerchner2026}. Its central claims may be summarised as follows. 

\begin{enumerate}
\item Computation is not an intrinsic physical process but a mapmaker-dependent abstraction: it is always computation-for-an-observer, imposed upon a physical system by an agent capable of constructing and applying an interpretive mapping.

\item Simulation of a process does not constitute instantiation of that process. The formal reproduction of the structure of a phenomenon is not the same as the realisation of the phenomenon itself; syntax does not entail or produce semantics or phenomenology.

\item Consciousness depends upon the actual physical realisation of certain kinds of dynamics. It is not portable across substrates in the way that pure information is portable: it is tied to specific modes of physical organisation.
\end{enumerate}

A key conceptual innovation is the introduction of the `mapmaker': an active, metabolically vulnerable agent that constructs computational descriptions. This concept is important because it makes explicit what is typically concealed in standard accounts of computation: the fact that computational descriptions presuppose choices that are not fixed by the physical facts alone. It is worth acknowledging that Miłkowski's mechanistic account attempts to circumvent this by grounding computational functions in natural selection: on his view, an evolved mechanism earns its computational description by virtue of the functional role it plays in the organism's survival, without requiring a Master Engineer or external interpreter.~\citep{Milkowski2013} This move is significant: evolutionary history does narrow the space of apt computational descriptions, and it provides a naturalistically grounded criterion that blocks the triviality objection. But it does not eliminate the residual underdetermination that the present argument identifies. A purely \emph{functional} description of an evolved mechanism---`this process contributes to survival by doing X'---may be naturalistically grounded without external interpretation. A specifically \emph{computational} description, however, requires something further: a choice of state-individuation, a specification of what counts as input and output, and a determination of which physical transitions correspond to which computational transitions. These choices are not fixed by the physical facts, nor by the evolutionary history, because the same physical mechanism with the same evolved functional role will support multiple incompatible computational descriptions under different state-individuation schemes---precisely the point established by Spencer et al.'s AND/OR argument.~\citep{Spencer2013} Natural selection selects for physical mechanisms that serve functional roles; it does not inscribe a unique computational description into those mechanisms. The 3M criterion of Kaplan and Craver provides principled guidance for identifying the most adequate computational description, but adequacy is an epistemic relation between a description and a mechanism, not an intrinsic property of the mechanism itself.~\citep{KaplanCraver2011} The mapmaker---whether a human engineer, an evolutionary theorist, or a cognitive scientist applying the 3M criterion---has not been eliminated by the mechanistic move; it has been given a richer and more constrained methodological context within which to operate. Computation does not happen in nature; it is described of nature, by agents equipped with the appropriate conceptual and methodological resources.

Importantly, the Lerchner framework, like Searle's, does not restrict consciousness to biological systems. It allows, in principle, that a non-biological system could realise the necessary dynamics, provided those dynamics are intrinsically organised rather than merely formally described. This is a significant concession to the spirit of computationalism, even as it rejects its letter: the question of machine consciousness is left open, but reframed as a question about physical implementation rather than formal description.

This reframing, moreover, resolves what might otherwise appear to be a residual difficulty in the simulation--instantiation distinction: the challenge of operationalising the criterion by which one determines, for an arbitrary physical system, on which side of the line it falls. The criterion is, in fact, neither arbitrary nor empirically intractable. It is the criterion of \emph{physical identity}. Instantiation just is identity of physical process: a storm instantiates a storm because it \emph{is} the relevant physical dynamics --- the pressure gradients, the convective cells, the moisture transport --- not a model or description of them. A weather simulation is not a storm because, however faithful its representation, it is a different physical process, constituted by voltage patterns in silicon, that is \emph{about} storms rather than being one. No increase in computational resolution closes this gap, because the gap is not one of degree but of categorical physical identity: the simulation and the storm are simply different things. The same logic applies without remainder to consciousness. The question of whether a given system instantiates consciousness is not a question about the sophistication or fidelity of its computational description; it is a question about whether that system \emph{is} the kind of intrinsically organised, autopoietically grounded physical process that grounds phenomenal experience, or whether it is a different physical process that models or represents such a process. A test for identity is, in principle, straightforwardly operationalisable: one determines whether the physical constitution of the candidate system is the relevant process itself, or merely a representation of it --- just as one can determine, without conceptual remainder, whether a map is the territory it depicts. Anything short of identity remains, at some level of description, simulation. This criterion is already implicit in everything the foregoing sections have established --- in Korzybski's insistence that the map is not the territory, in Searle's observation that a storm simulation does not make anyone wet, in Spencer's demonstration that computational descriptions are observer-assigned rather than intrinsic, and in the autopoietic account of living systems as constitutively self-producing rather than externally described. The identity criterion unifies these arguments into a single, decisive principle: what the machine consciousness thesis must demonstrate is not that a computational system models consciousness with sufficient fidelity, but that it \emph{is} the relevant physical process. That demonstration has not been made, and the reasons canvassed in this paper give strong grounds for doubting that it can be.

A mechanistic computationalist might respond that identifying metabolic, autopoietic organisation as the relevant physical process is simply a more specific determination of which mechanisms must be present---and that specifying the required mechanisms more precisely is entirely consistent with MC. Individual metabolic reactions are, after all, mechanistic in the standard sense: enzyme kinetics, phosphorylation cascades, ion channel dynamics are all causally organised, physically realised processes of exactly the kind MC requires. On this reading, the present paper would simply be telling MC where to look, not challenging its framework.

The response fails, however, because what the identity criterion requires is not the right individual mechanisms but the right \emph{organisational closure}---and closure to efficient causation, in Rosen's sense, has a causal structure that is categorically irreducible to any sum or arrangement of mechanisms.~\citep{Rosen1985} Rosen's central argument in his relational biology is that organisms are not mechanisms: a mechanism is a system whose organisation can be fully captured by efficient causation alone, with the organisation imposed from without by a designer or external context. A living system, by contrast, exhibits closure to efficient causation---each component of the metabolic network is produced and maintained by the activity of the whole, and the whole is causally efficacious with respect to its own parts in a way that no mechanism can replicate. This is not a more specific specification of which mechanisms must be instantiated; it is a categorically different kind of causal organisation. The individual metabolic reactions may each be mechanistic; their closure is not. It is the closure that is constitutive of life, and it is the closure that no computational or mechanistic description can fully capture---not because we lack the right description, but because the closure is precisely what makes the system the kind of thing it is, rather than a collection of mechanisms that a description organises from without. The MC counter-move therefore dissolves: asking for the right mechanisms is asking for the parts; the identity criterion requires the whole, and the whole, in the relevant sense, is not the sum of its mechanistic parts. This is not merely a philosophical assertion: Rosen established it mathematically using the tools of category theory in his analysis of $(M,R)$-systems, and his former student Aloisius Louie subsequently formalised and completed Rosen's proofs using finite category theory and lattice theory, replacing Rosen's looser use of Russell's vicious circles with rigorously defined hierarchical cycles and clef systems.~\citep{Rosen1985, Louie2013} The resulting proof establishes not only that anticipatory systems cannot be fractionated without losing their anticipatory properties, but that closure to efficient causation is an \emph{all-or-nothing} algebraic property: a system either possesses a fully closed self-entailing mapping network or it does not. There is no partial closure, no system that is `more or less' alive in the relevant organisational sense. This entails a topological discontinuity between the class of complex mechanisms and the class of living systems that cannot be bridged by any gradual, linear accumulation of mechanical components. As Rosen and Louie both noted explicitly, every individual physical part or isolated process within an organism can be described completely by a mechanism; the non-reducibility applies only to the totality of the organisational network. The mathematical structure is precisely analogous to a manifold and its atlas of local flat approximations: at every point the surface can be mapped by a locally flat chart, and a complete atlas of such charts can be constructed, yet the manifold is not flat, and no assembly of flat charts entails that it is. The local approximation is indispensable; the global inference is unwarranted. Crucially, a piecewise linear function can approximate any continuous nonlinear function arbitrarily closely by making the linear segments shorter and shorter, yet the function never becomes linear; analogously, adding mechanical components to a system increases its complexity without limit but never produces closure to efficient causation, because the transition is discontinuous rather than continuous. Rosen's category-theoretic result establishes the same structure for living systems: a successful scientific programme of mechanistic description of all the components of a living system does not compel the conclusion that the anticipatory system as a whole is reducible to mechanism. The irreducibility of organisational closure is not a gap in our knowledge but a proven mathematical feature of the class of systems to which living organisms belong.

A sophisticated recent challenge to this conclusion is posed by Chrisley, who argues that rejecting computational functionalism---the claim that computation alone suffices for consciousness---does not establish that machine consciousness is impossible.~\citep{Chrisley2026} The logical point is valid: from `C is not sufficient for P' it does not follow that `it is impossible for C and P to occur together'. Chrisley introduces the notion of an INS condition---an Insufficient but Necessary part of a Sufficient condition---to argue that computation might be a non-redundant component of some broader sufficient basis for consciousness, even if insufficient on its own. He distinguishes two senses of conscious AI: Type-2, in which a system is conscious \emph{solely} in virtue of implementing a formal digital computation (which requires computational functionalism and is what our paper refutes), and Type-1, in which computation plays a major constitutive or enabling role within a broader sufficient basis for consciousness (which does not require computational functionalism and which Chrisley suggests is the more plausible and neglected scenario). The present paper's argument is directed at Type-2; it does not, as stated, rule out Type-1.

However, when the Type-1 scenario is examined carefully in the light of the observer-relativity argument, it reduces to a claim that is entirely compatible with our paper's thesis and that constitutes no independent challenge to it. To say that computation plays an enabling role in consciousness is to say that some of the physical processes involved in the broader sufficient basis for consciousness can, if one were so inclined, be described at a computational level of abstraction. The enabling is done entirely by those physical processes; the computational description is a potentially useful but ontologically idle characterisation of what they are doing. Observer-relativity attaches to the description, not to the underlying physical processes the description picks out. This means that `computation plays an enabling role' reduces, on analysis, to `some physical processes that happen to be describable computationally participate in the relevant physical dynamics'---which is true of biological neural systems and is entirely consistent with metabolic physicalism. Chrisley's McLaren analogy, pressed to this conclusion, supports rather than threatens our position.\footnote{Chrisley introduces the McLaren analogy to illustrate the logical structure of his INS argument.~\citep{Chrisley2026} The analogy runs as follows. Suppose someone claims that \emph{driving a McLaren is sufficient for winning a Formula One Grand Prix}---call this `McLarenism'. McLarenism is false: driving a McLaren is not enough to win. One also needs driving skill, a well-functioning car, a race strategy, good pit-crew execution, and favourable conditions. The falsity of McLarenism does not, however, show that a McLaren \emph{cannot} win a Grand Prix, nor that the car is irrelevant to victory. It shows only that the car is not sufficient on its own. The McLaren may nonetheless be a non-redundant component of one particular sufficient package for winning: remove the car, and that specific sufficient basis no longer obtains, even if other sufficient bases (involving Ferrari, Mercedes, or another team) remain possible. By analogy, the falsity of computational functionalism---the claim that computation alone suffices for consciousness---does not show that a computational system cannot be conscious, nor that computation is irrelevant to consciousness. Computation may still be a non-redundant component of some broader sufficient basis for consciousness that also includes embodiment, metabolic organisation, or other non-computational elements. Chrisley uses the analogy to argue that the debate should shift from `Is computation sufficient for consciousness?' to `Can computation be a non-redundant component within some sufficient basis for consciousness?'---what he calls an INS (Insufficient but Necessary part of a Sufficient) condition.} the race is won by the physical car, driver, and team, not by the description `McLaren'; saying computation plays an enabling role is like saying `the McLaren plays an enabling role'---true in the sense that we can apply that label to part of the winning package, but the label does not pick out anything over and above the physical processes that constitute winning. The observer-relativity argument therefore applies to Type-1 as well as Type-2, once Type-1 is examined carefully: in both cases, the computational description is applied to physical processes from without, and those physical processes do all the real work. What Chrisley helpfully clarifies, and what our paper's positive account addresses, is the question of what those physical processes must be---and our answer is that they must be metabolically grounded, autopoietically organised, and constituted by the identity of the relevant dynamics, not by any computational description of them. Chrisley and Shardlow are right that this `missing middle' scenario---consciousness depending on non-computational physical organisation but in principle realisable in non-biological substrates---is the most plausible route to artificial consciousness and the one most neglected in the debate; and metabolic physicalism, as developed in this paper, is precisely an account of what that scenario requires.~\citep{Shardlow2026} It is worth adding that Louie's proof of the all-or-nothing character of closure to efficient causation provides formal grounds for rejecting any gradualist version of the machine consciousness argument---including gradualist readings of the INS framing.~\citep{Louie2013} The suggestion that adding sufficiently many computational or mechanistic components to a system will eventually yield the sufficient basis for consciousness assumes that the transition is continuous and can be reached incrementally. Louie's proof establishes that it is not: there is a topological discontinuity between the class of mechanisms and the class of systems closed to efficient causation that no incremental addition of components can bridge. A gradualist INS argument is therefore formally analogous to arguing by induction that stacking bricks can reach the Moon: the inductive step is valid at each stage, but the conclusion does not follow because the target condition is discontinuously beyond the reach of the method.

A sophisticated recent challenge to this line of argument is posed by what Kanai has termed \emph{intrinsic computational functionalism} (ICF).~\citep{Kanai2026} Kanai draws a careful distinction between observer-relative semantic interpretation---the labels assigned to physical states, the calling of a voltage `1' or `0', the attribution of symbolic meaning to a gate transition---and intrinsic causal organisation: the actual physical causal structure of the system, in which voltages constrain voltages, memory states transition, and recurrent dynamics unfold, independently of any observer. His slogan captures the distinction neatly: `A map needs a mapmaker. A mechanism does not.' ICF then proposes that consciousness depends on intrinsic physically realised causal, dynamical, integrated, and geometric organisation---not on externally imposed semantic interpretation. This is a more careful and more defensible position than classical computational functionalism, and it deserves a direct response.

The response has two parts. The first concerns the multiple realisability argument already developed above. Kanai is right that the causal organisation of a physical system is intrinsic and observer-independent. Voltages constraining voltages is a physical fact regardless of how it is described. But the criterion `the right intrinsic causal, dynamical, integrated, and geometric organisation' is a type-level description: it picks out an equivalence class of physical systems that share the relevant organisational properties. Any such equivalence class is, as we have argued, an abstract type, however physically constrained. The `intrinsic' in ICF refers to the physical reality of each token; but `the right organisation' is a type, and types are abstract even when tokens are concrete. ICF therefore does not escape the multiple-realisability-equals-abstraction problem; it merely relocates the abstraction from semantic labels to causal structure. A mapmaker is still needed to identify which physical systems fall within the relevant equivalence class---not to assign semantic labels, but to determine which causal-dynamical-geometric profiles count as the right ones.

The second part of the response is more fundamental. ICF specifies the relevant physical organisation in functional terms: what the system \emph{does} causally, the pattern of its state transitions, the geometry of its dynamical manifold. But what a system does causally and what it physically \emph{is} can come apart in precisely the way that matters for consciousness. Two systems can share the same causal-dynamical-geometric profile---the same functional organisation in Kanai's sense---without sharing the same physical constitution. The positive account advanced in this paper insists that the relevant physical constitution for consciousness is not merely causal-dynamical but metabolic and autopoietic: a system invested in its own continued existence through self-producing, far-from-thermodynamic-equilibrium organisation possesses something that a system with the same causal-dynamical profile but no metabolic grounding lacks---intrinsic normativity, genuine agency, a real stake in its own future. ICF has no resources to capture this distinction because it specifies the relevant organisation in purely functional-causal terms, which are silent on the question of whether the system has anything genuinely at stake. A thermostat and a neuron both involve voltage changes causing further state transitions; their causal organisation overlaps substantially; neither is conscious in virtue of that overlap. The identity criterion, applied to Kanai's own framework, asks not whether the system has the right causal organisation but whether it physically \emph{is} the process that constitutes consciousness---and that process, on the present account, is autopoietically grounded metabolic organisation, not causal-dynamical structure per se, however intrinsic.

\section{Metabolism and the Physical Basis of Consciousness} \label{Metabolism}
A central theme emerging from the synthesis developed in this paper is the role of metabolism and self-maintenance in grounding consciousness. Both the Lerchner framework and the work of Nasuto and Bishop converge on the claim that consciousness is associated with systems that are self-organising, metabolically active, or functionally equivalent to metabolically active systems, and that are intrinsically motivated by internal states.~\citep{Lerchner2026, NasutoBishop2011}

The most formally rigorous account of what distinguishes such systems from mechanisms and computational processes is provided by Robert Rosen's relational biology, and in particular his theory of \emph{anticipatory systems} and his \emph{metabolism-repair} ($(M,R)$) systems.~\citep{Rosen1985} Rosen's central claim is that living systems are distinguished from all mechanisms by the presence of \emph{closure to efficient causation}: in a living system, each component is produced and maintained by the causal activity of the whole, and the whole is causally efficacious with respect to its own parts in a way that no mechanism---no system whose organisation is imposed from without---can replicate. A mechanism has its organisation externally provided, whether by an engineer or by an evolutionary history that selects for functional roles; a living system produces and sustains its own organisation through its own activity. From this formal structure, Rosen derives a further property that is central to the present paper's argument: living systems are irreducibly \emph{anticipatory}. Because they contain internal models of their own future states and of the environment, they act in advance of events rather than merely in response to them, and their present behaviour is partly constituted by the model's predictions about what is to come. It is important, however, to note a distinction that Rosen's work points towards but that was formalised precisely by his former student, the mathematical biologist Aloisius Louie: \emph{anticipatory systems} form a broader class than \emph{living systems}.~\citep{Louie2013} An economy, a corporate strategy, or an active-inference AI network may possess predictive models and qualify as anticipatory without being alive, because their predictive models are built and maintained by external agents rather than by the system's own internal (M,R) loop. The distinctively biological is not anticipation alone but anticipation together with full closure to efficient causation: a living system, and only a living system, manufactures and repairs its own predictive model from within its own metabolism. This distinction is directly relevant to the machine consciousness thesis: even granting that a sufficiently sophisticated AI system might qualify as genuinely anticipatory in the relevant sense, this would not suffice to make it a living system or a candidate for consciousness, because it would still require external agents to build and maintain its predictive architecture. As established in the previous section, Louie --- formalising and completing Rosen's proofs using finite category theory and lattice theory --- proved that organisational closure cannot be reached by gradual accumulation of mechanical parts. Closure to efficient causation is an all-or-nothing algebraic property: a system either possesses a fully closed self-entailing mapping network or it does not. There is a topological discontinuity between a highly complex open mechanical system and a closed living relational system that no incremental addition of components can bridge. The irreducibility of living organisation to mechanism is therefore not a philosophical conjecture but a proven mathematical result.

Biological organisms are not passive processors of information: they are systems that are \emph{invested} in their own continued existence, that have something at stake in the outcomes of their interactions with the environment, and whose internal states carry genuine normative significance in virtue of that investment. Hunger is not merely a computational state that causes food-seeking behaviour; it is a condition of a living body that \emph{matters} to that body. The mattering is not imposed from without by an interpreter; it arises from within, from the metabolic structure of the organism itself. A position discussed at length by Evan Thompson in his 2007 book, `Mind in Life'.~\citep{Thompson2007}

Two further theoretical accounts complement Rosen's framework by supplying what it formally presupposes but does not itself develop in biological or philosophical detail. The first is Maturana and Varela's theory of \emph{autopoiesis}, which provides the biological instantiation of Rosen's closure concept.\footnote{The term \emph{autopoiesis} (from the Greek \textit{autos}, self, and \textit{poiesis}, making or creation) was introduced by the Chilean biologists Humberto Maturana and Francisco Varela to characterise the defining property of living systems.~\citep{MaturanaVarela1980} An autopoietic system is one that continuously produces and maintains itself by generating its own constituent components through its own internal processes: the system's organisation is the product of the system's own activity, rather than being imposed or sustained from without. A living cell is the paradigm case---it does not merely process inputs and produce outputs, but actively regenerates the very molecular machinery that enables it to do so. By contrast, an \emph{allopoietic} system (a clock, a robot, a computer) is one whose components are produced and organised by an external agent: the system produces something other than itself. The distinction is thus not merely one of complexity or adaptive sophistication, but of the fundamental locus of organisation---whether it originates from within the system or is imposed upon it from without. In Rosen's terms, autopoiesis is the biological realisation of closure to efficient causation. Letelier, Mar\'in, and Mpodozis subsequently established formally that autopoiesis is a highly restricted, materialised special case of an $(M,R)$ system, occupying the innermost position in the entailment hierarchy: impredicative systems $\supset$ anticipatory systems $\supset$ $(M,R)$ systems $\supset$ autopoietic systems.~\citep{LetElierEtAl2003} Cornish-Bowden and colleagues further translated the abstract $(M,R)$ network into real biochemical equations --- interlocking catalytic cycles and enzyme degradation pathways --- demonstrating that Rosen's and Louie's abstract relational constructs are not biologically idle but mirror actual metabolic pathways.~\citep{CornishBowden2007}} Autopoiesis specifies concretely what Rosen's closure means in biochemical terms: the living cell continuously regenerates the molecular machinery that constitutes it, each component produced by the metabolic activity of the whole. The second complementary account is Mark Bickhard's \emph{interactivist} model of mind, which provides the normative bridge between Rosen's formal structure and the paper's claims about intrinsic normativity and consciousness.~\citep{Bickhard2009} Rosen tells us what kind of system an anticipatory system is; Bickhard tells us why such systems have genuine normative properties rather than merely functional ones. Bickhard argues that the dominant tradition in cognitive science commits what he calls the `encodingist' error: the assumption that mental representation consists in the possession of internal states that encode, or stand in for, features of the environment in the manner of a code or picture.\footnote{Bickhard's critique of the encodingist error converges in instructive ways with Roy Harris's integrational rejection of what Harris terms the `language myth': the orthodox view that communication consists in the telementation of pre-formed meanings from sender to receiver via a shared fixed code.~\citep{Harris1981} Harris argues that this picture fundamentally misconstrues the nature of linguistic meaning, which is not a determinate content transmitted between minds but arises through the contextually situated, temporally embedded activity of communicating agents. On Harris's integrational account, signs are not pre-existing vehicles for meaning; they are constituted in situ through the interpretative practices of participants. The parallel with Bickhard is precise: just as Bickhard shows that the encodingist picture presupposes the very semantic relations---the correspondence between inner states and worldly features---that it is invoked to explain, Harris shows that the code model presupposes a determinacy of meaning that communication is, in fact, engaged in constructing. In both cases, the underlying error is to treat meaning as intrinsic to representations or signals, rather than as emergent from the normative, context-sensitive interactions of situated agents. Together, these critiques reinforce the broader claim of this paper: that neither internal encoding nor external transmission suffices to explain the origin of semantics, and that any account of meaning adequate to ground consciousness must locate normativity in the dynamics of agents genuinely invested in their own continued existence, rather than in the formal structure of the representations they manipulate.} This assumption renders the normativity of representation inexplicable---it cannot account for the possibility of \emph{misrepresentation}, since no intrinsic property of a symbol determines whether it is correctly or incorrectly applied. The remedy requires a shift from a substance-based to a \emph{process}-based metaphysics. Together, the three accounts form a coherent and mutually supporting positive picture: Rosen provides the mathematical backbone, autopoiesis provides the biological detail, and Bickhard provides the philosophical bridge to normativity and consciousness. No purely computational process can replicate this structure, because computation is observer-assigned and therefore cannot generate the intrinsic normativity that Bickhard identifies, realise the organisational closure that Rosen proves is irreducible, or instantiate the self-producing biological organisation that Maturana and Varela describe.

Striking recent empirical support for this account comes from an unexpected direction. Rajan et al.\ have demonstrated that the single-celled organism \emph{Stentor coeruleus}---a ciliate with no nervous system, no synapses, and no brain---exhibits genuine habituation: a form of non-associative learning in which repeated mechanical stimulation produces a graded and reversible reduction in contraction probability.~\citep{RajanEtAl2026} The molecular mechanisms responsible have now been characterised in considerable detail: habituation in \emph{Stentor} is mediated by calcium signalling, calcium/calmodulin-dependent protein kinase II (CaMKII) activity, and the phosphorylation and dephosphorylation of cell-surface mechanoreceptors, in a manner that closely parallels the synaptic plasticity mechanisms underlying learning in metazoan nervous systems. Memory of the habituated state can be propagated to daughter cells across cell division. What is philosophically decisive for the present argument is not merely the fact of single-cell learning, but the \emph{mechanism} by which it is achieved. The learning is not described by these molecular dynamics; it \emph{is} them. There is no computational architecture, no observer-assigned symbol system, no abstract programme being executed: there is only the intrinsic physical chemistry of a metabolically active, autopoietically organised cell modifying its own receptor proteins through calcium-dependent kinase cascades in response to experience. This is metabolic physicalism made empirically concrete. The \emph{Stentor} case also provides a precise illustration of the identity criterion developed above: the question of whether this organism learns is settled not by asking what computation its molecular dynamics implement, but by determining what those dynamics physically \emph{are}---and they are, it turns out, the very processes that constitute learning at the cellular level in this organism, even if other metabolic and molecular processes serve analogous constitutive roles in other organisms. Recent work in microbial cognition more broadly confirms the importance of distinguishing between genuine experience-driven learning of this kind and behaviourally similar adaptive responses that are driven by evolutionary processes rather than individual experience---a distinction that maps directly onto the philosophical difference between intrinsic normativity and observer-assigned significance that this paper has been at pains to draw.~\citep{MesserOnaKost2026}

This suggests that consciousness arises in systems that are physically invested in their own continued existence---systems for which there is, in a robust and non-derivative sense, something at stake. Such systems possess an intrinsic normativity wholly absent from purely computational processes: values that are theirs, not assigned to them.

On Bickhard's account, representation is not a static structural relation between an inner state and an outer object but an emergent property of a system's dynamic interaction with its environment: specifically, the system's \textit{anticipation} of further possible interactions, grounded in its ongoing activity of self-maintenance far from thermodynamic equilibrium.\footnote{The idea that anticipation is constitutive of intelligent behaviour rather than auxiliary to it has deep roots in the history of cybernetics. In their seminal 1943 analysis of purposive behaviour, Arturo Rosenblueth, Norbert Wiener, and Julian Bigelow observed that a cat pursuing a running mouse does not run towards the mouse's current position but towards an extrapolated future one.~\citep{RosenbluthWienerBigelow1943} The insight is deceptively simple yet profound: adaptive agents act not on the world as it is, but on the world as they expect it to become. Prediction is therefore not a supplementary function layered atop perception; it is constitutive of purposive behaviour itself. This cybernetic observation finds its fullest theoretical elaboration in the framework developed by the mathematical biologist Robert Rosen in \emph{Anticipatory Systems}.~\citep{Rosen1985} Rosen formalises the distinction the cyberneticians had identified: a purely reactive system---a thermostat, a servo-motor, a classical computational architecture---responds to what is happening now, driven exclusively by its current state and its history. An anticipatory system, by contrast, carries within itself an internal model of its own future interactions with the environment and uses that model to modify its current behaviour in advance of events rather than merely in response to them. The model is not a passive representation but a causally active component of the system's physical organisation---what distinguishes, in Rosen's terms, a system that merely reacts from one that genuinely \emph{acts}. Crucially for the argument of this paper, Rosen insists that anticipatory systems cannot be adequately described within the framework of classical mechanics or computation theory, both of which are restricted to reactive, state-determined dynamics in which the future can play no causal role in the present. Living organisms escape this restriction because their internal models---themselves physical structures, products of evolution and development---encode possible futures and feed those encodings back into present dynamics. The parallel with Bickhard is then precise. Bickhard locates representation in the system's anticipation of further possible interactions grounded in metabolic self-maintenance; Rosen locates the distinctively biological in the possession of an internal predictive model that shapes present action. Both trace this capacity to the same source the cyberneticians first identified: the organism's constitutive orientation towards its own future. All three accounts converge on a conclusion directly relevant to the abstraction fallacy: genuine anticipation, and the normativity it makes possible, cannot be captured by any purely reactive, input--output computational description, however complex, and presupposes a physically grounded system that is invested, in the relevant sense, in what happens next. It is worth noting, in this context, that the Good Regulator Theorem of Conant and Ashby---which states that every good regulator of a system must contain a model of that system---is sometimes invoked in support of the machine consciousness thesis, on the grounds that any system sophisticated enough to regulate its environment effectively must contain an internal model, and that this model-containment is sufficient for some form of cognition or experience.~\citep{ConantAshby1970} The invocation does not succeed. The Good Regulator Theorem is a result about control systems: it establishes a necessary condition for effective regulation, not a sufficient condition for consciousness. A thermostat contains a model of room temperature in the relevant sense and is an effective regulator of it; no one concludes it is conscious. More fundamentally, the theorem establishes that the model must be present, not that the model must be the kind of causally active, metabolically grounded, anticipatory structure that Rosen, Bickhard, and the autopoietic account identify as constitutive of genuine agency and normativity. The Good Regulator Theorem is, in effect, a result about the necessity of internal models for regulation---which our paper grants entirely---without touching the question of what kind of physical organisation those models must be embedded in if they are to ground phenomenal experience.} Normativity---the capacity for a representation to be correct or incorrect---emerges from this thermodynamic grounding: a representation is correct if acting upon it serves to maintain the system's far-from-equilibrium organisation, and incorrect if it does not.

Crucially, this account of normativity is \emph{intrinsic} to the system: it does not depend upon the interpretive activity of an external observer, but arises from the system's own metabolic investment in its continued existence. In this respect, Bickhard's interactivism provides precisely the theoretical underpinning that the metabolic physicalism advanced here requires: it explains, at the level of process metaphysics, why the intrinsic normativity that consciousness presupposes cannot be derived from static computational structure, and why it is a characteristic of systems that realise the anticipatory, self-maintaining organisation Rosen identifies.

It is important to situate the identity criterion developed in this paper precisely within the landscape of identity theories of mind. The criterion is a claim about \emph{token} identity, not \emph{type} identity. Type identity theory — the view associated with Place and Smart, and subsequently targeted by Putnam's multiple realisability argument — holds that mental state \emph{types} are identical to physical state types: pain-as-such is identical to C-fibre-firing-as-such. Our paper makes no such claim. We do not assert that consciousness as a type is identical to any particular physical type such as metabolic autopoietic organisation; we assert that each \emph{instance} of consciousness is identical to some specific physical process, and that the question of whether a given system is conscious is settled by determining whether its physical constitution \emph{is} the relevant process rather than merely modelling or resembling it. Token identity theory is largely immune to the multiple realisability objection in its classical form: the fact that pain can be realised in physically different systems — humans, octopuses, silicon, in principle --- does not threaten the claim that each particular pain event is identical to a particular physical event in the system that has it. Our paper is therefore compatible with a suitably restricted form of multiple realisability: different physical processes may constitute consciousness in different organisms or systems, provided each is genuinely the relevant process in that system and not merely a computational description of it.

A related objection, pressing harder on the identity claim, derives from Leibniz's Law: if a mental state and a physical state are identical, they must share all properties; but mental states have phenomenal, intentional, and normative properties that physical states apparently lack, whilst physical states have spatial, electrochemical, and thermodynamic properties that mental states apparently lack. The objection has force against naive formulations of identity theory, but it does not threaten the present account, for the following reason. The physical processes we identify as constituting consciousness---autopoietically organised, metabolically grounded, far-from-equilibrium dynamics---are not bare electrochemical events to which intentional and normative properties are subsequently added by an observer. On Bickhard's interactivist account, intrinsic normativity \emph{arises from} thermodynamic self-maintenance: a representation is correct or incorrect in virtue of whether acting on it serves to maintain the system's far-from-equilibrium organisation, and this correctness condition is a property of the physical process itself, not a property projected onto it from without. Similarly, on the autopoietic account, the self-producing metabolic organisation of a living system constitutes a genuine biological perspective---an intrinsic orientation towards the system's own continued existence---which is not observer-assigned but physically grounded. The phenomenal, intentional, and normative properties of consciousness are, on this view, intrinsic properties of the relevant physical organisation, not additions to it; Leibniz's Law is therefore satisfied rather than violated. What the critic is implicitly assuming is that the physical description must be given at the level of electrochemical microstates, stripped of all organisational properties; but this is to beg the question against any account that locates the relevant physical organisation at a higher level of physical description, as both token identity theory and metabolic physicalism do.\footnote{A related and widely discussed objection to physicalist accounts of consciousness is Frank Jackson's knowledge argument, in which Mary the colour scientist knows all physical facts about colour vision yet learns something new---what it is like to see red---upon first leaving her black-and-white room.~\citep{Jackson1982} The standard gloss is that this shows phenomenal facts to be non-physical. Construed carefully, however, the knowledge argument does not threaten our paper's position; it \emph{supports} it. What Mary possesses before leaving the room is precisely a complete functional-computational description of colour vision: every wavelength, every retinal response, every neural firing pattern, every causal-informational transition in the colour processing pathway. She has, in Lerchner's terms, the most detailed and faithful \emph{map} of colour experience that could be drawn. And the central claim of our paper is that possessing the map, however perfect, is not the same as instantiating the territory. The phenomenal residue---what she learns on seeing red---is exactly what our paper predicts would be missing from any purely computational description, however complete. Far from being a counterexample to metabolic physicalism, Mary's situation is a precise illustration of the simulation--instantiation distinction: complete computational-functional knowledge does not constitute phenomenal instantiation. It is also worth noting that Jackson himself later retreated from the argument, adopting what is known as the \emph{phenomenal concept strategy}: the view that phenomenal concepts refer directly to physical states without passing through functional descriptions, which explains why Mary acquires a new \emph{concept} without there being a new \emph{fact}.~\citep{JacksonLater2003} This later position is, in fact, congenial to metabolic physicalism: if phenomenal concepts refer directly to physical states rather than via functional descriptions, then it is instantiation of the physical process---not computational modelling of it---that grounds phenomenal experience. Jackson's retreat therefore reinforces rather than undermines the present paper's argument.}

The positive account that emerges from this synthesis may be described as a form of \emph{metabolic physicalism}: the view that consciousness is grounded in self-maintaining, internally organised physical dynamics, and that the formal structure of computation, whilst a useful tool for describing and predicting behaviour, is neither necessary nor sufficient for the instantiation of mind.

\section{The Mind--Body Problem: How Consciousness May Exert Causal Power} \label{MindBody}
The mind--body problem concerns the relationship between subjective, first-person experience and objective physical processes. In contemporary debate, this issue is frequently reframed as the question of whether consciousness can be fully explained in terms of abstract computational structure, or whether it depends essentially on the nature of its physical realisation. The analysis developed in this paper supports the positive thesis advanced by Searle and Lerchner: whilst rejecting computationalism, it retains a broadly physicalist position in which consciousness depends on the actual instantiation of specific kinds of intrinsic physical dynamics, rather than on formal symbol manipulation.

On this account, the mind--body problem is not resolved by identifying the appropriate computation but by identifying the appropriately organised physical processes---potentially including metabolically grounded, self-maintaining dynamics. Crucially, such processes need not be confined to biological systems. In principle, a non-biological system engineered to realise the same relevant dynamics could also instantiate conscious experience, thereby reframing the mind--body problem as a question concerning the conditions under which physical systems---biological or synthetic---can generate genuine phenomenology.

However, if consciousness is an emergent feature of intrinsically organised physical systems invested in their own continued existence, as the foregoing account proposes, a further and deeply important question arises: is consciousness epiphenomenal, or does it exert genuine causal power? The classical formulation of the mind--body problem foregrounds the difficulty of assigning a scientifically tractable causal role to subjective experience.

Various responses have been proposed, of which a very small selection is sketched below:

\begin{enumerate}
\item An early and influential attempt was the synaptic indeterminacy model developed by Popper and Eccles in \emph{The Self and Its Brain} and elaborated in Eccles's later collaboration with the physicist Friedrich Beck.~\citep{PopperEccles1977, BeckEccles1992} The model begins from an empirical observation that Eccles, as a Nobel laureate in neurophysiology, was unusually well placed to press: the release of neurotransmitter vesicles at the synaptic cleft is an irreducibly probabilistic process. The pre-synaptic terminal does not deterministically discharge its vesicles; rather, it presents, at any given moment, a quantum-mechanical probability distribution over possible discharge events. On the standard physical picture, which outcome actually occurs is a matter of chance. Eccles argued that this microscopic indeterminacy furnishes precisely the opening that a non-physical mind requires in order to act upon the brain without violating the conservation of energy: the mind does not \emph{inject} energy into the system, but \emph{selects} among outcomes that physics already renders possible.

In the later Beck--Eccles formulation, this selective intervention was modelled as quantum tunnelling: psychons---postulated mental units, each reciprocally linked to a cortical `dendron' (a cylindrical bundle of approximately one hundred pyramidal neurons in the neocortex, constituting the fundamental anatomical unit of conscious processing)---act upon the pre-synaptic membrane to modulate the probability of exocytosis, thereby biasing neuronal firing in the service of voluntary action and conscious intention.

%The causal structure of the proposal is thus carefully constrained: physics supplies a probability distribution over physically possible neural outcomes; mind, operating through psychon--dendron coupling, shifts the weights of that distribution without overriding it. The model is explicitly interactionist rather than epiphenomenalist, and it was designed to show that mental causation need not conflict with physical law, provided that physical law is understood in its genuinely indeterministic, quantum-mechanical form rather than in the classical deterministic guise that rendered Cartesian interactionism seemingly incoherent.

%Critics have objected that the proposed mechanism remains underspecified---the ontological status of psychons is nowhere made rigorously precise---and that thermal noise at physiological temperatures may be sufficient to swamp any quantum-coherent effect at the synaptic level, a concern that applies with equal force to the Penrose--Hameroff programme discussed below. Nevertheless, the Popper--Eccles model retains considerable philosophical interest, both as a serious attempt to ground mental causation in the microphysics of the brain, and as an early recognition that the explanatory gap between consciousness and its physical substrate cannot be closed by invoking the classical, deterministic image of neural computation alone.

\item The `Orch-OR' model advanced by Penrose and Hameroff posits that quantum processes in neuronal microtubules account for conscious causation, grounding the non-algorithmic character of mathematical insight in quantum-gravitational state reduction~\citep{Penrose1994, HameroffPenrose2014}. This proposal remains highly contested within the neuroscientific and philosophical communities, but it has the considerable merit of taking seriously the demand for a physical mechanism through which consciousness could do causal work.

\item Most recently, Tom Froese has proposed `Irruption Theory', according to which conscious experience may exert causal influence in contexts of high entropy and systemic under-determination, where multiple future trajectories are physically permissible~\citep{Froese2023}. On this view, consciousness is not merely an inert by-product of physical processes but may actively shape their unfolding under specific conditions---conditions that characteristically obtain at moments of genuine choice, creative thought, and novel action. It is worth noting that Irruption Theory addresses a question distinct from, and complementary to, that addressed by Rosen's anticipatory systems framework elsewhere in this paper. Rosen asks what kind of physical organisation a system must have in order to be genuinely anticipatory and agentive---his answer being closure to efficient causation and internal predictive modelling. Froese asks how, given that a system is conscious, its consciousness might do real causal work in the physical world---his answer being that it acts at moments of physical underdetermination where the future is genuinely open. The two accounts are therefore not in competition: Rosen specifies the conditions under which consciousness is possible; Froese proposes a mechanism by which it might be causally efficacious. Together they complement the present paper's positive account, grounding consciousness in metabolic organisation whilst preserving the possibility that it exerts genuine rather than epiphenomenal causal power.
\end{enumerate}

All these proposals resist the epiphenomenalist conclusion and insist on the genuine efficacy of the mind. Whether any of these succeed in their specific mechanistic claims is a matter for ongoing empirical and philosophical investigation. What the present analysis establishes is that a non-computational account of mind---one grounded in intrinsically organised physical dynamics---is at least as well placed to address the problem of mental causation as its computationalist rival, and arguably better placed, since it does not face the additional difficulty of explaining how an abstract formal structure could possess causal powers of any kind.

\section{Conclusion} \label{Conclusion}
The critique of computationalism advanced by the Lerchner paper is not novel in its core insights. It represents, rather, a synthesis and formalisation of a long-standing and multi-authored tradition of objection---one that has gathered force and precision over several decades without yet having produced the decisive shift in the field that its proponents believe it warrants.

From Korzybski's distinction between map and territory~\citep{Korzybski1933}, through Searle's demonstration that syntax is insufficient for semantics~\citep{Searle1980}, Floridi's analysis of the methodological status of computational description~\citep{Floridi2011}, Spencer's formal argument for the observer-relativity of computation~\citep{Spencer2013}, Winograd and Flores's grounding of computational meaning in social practice~\citep{WinogradFlores1986}, and the biologically oriented arguments of Nasuto and Bishop~\citep{NasutoBishop2011}---a consistent and convergent picture emerges.

Computation is an abstract, observer-dependent description. Consciousness, by contrast, is an intrinsic feature of certain kinds of physically organised systems: systems that are self-maintaining, metabolically grounded, and genuinely invested in their own continued existence. The conflation of these two levels---Lerchner's `abstraction fallacy'~\citep{Lerchner2026}---has generated persistent confusion in the philosophy of mind and artificial intelligence, and has encouraged unwarranted optimism about the prospects for machine consciousness on the basis of purely formal criteria.

The viable alternative is a form of non-computational physicalism in which consciousness depends upon intrinsically organised, self-maintaining physical dynamics. Such a framework preserves the genuine possibility of artificial consciousness---it would be dogmatic to exclude it---whilst firmly rejecting the sufficiency of computation alone. The question of whether a given physical system is conscious is not answered by determining what computation it implements, but by determining what kind of thing, physically and intrinsically, it \emph{is}.

It is worth acknowledging, finally, a broader observation about the sociology of this debate that bears directly on how the paper's arguments should be received. As recent exchanges in the literature make plain, whether machine consciousness is possible has, for many participants, acquired the character of a prior commitment rather than a conclusion reached through argument: positions are adopted first and defended post hoc by arguments fashioned to fit them. This is not a charge levelled at any individual interlocutor, but a structural observation about a field in which the central variable---consciousness itself---cannot be directly measured, making every argument underdetermined by the evidence available to it. Ganesh has shown formally that this underdetermination is not accidental: because P\_C (the computational condition) and $\Omega$ (the physical constraint condition) are perfectly collinear across all data in which consciousness can be confidently attributed, and because consciousness itself cannot be observed directly, any observer applying rigorous inductive methods will converge on computational functionalism regardless of whether it is true.~\citep{Ganesh2026} CF is, on this analysis, the inevitable output of the epistemic position its proponents occupy---not primarily the product of evidence. This structural diagnosis does not settle the first-order question; but it does change what a genuine defence of the machine consciousness thesis would have to look like. It is no longer sufficient to show that computation is \emph{necessary} for consciousness, or that the evidence is consistent with sufficiency: the CF proponent must supply evidence for sufficiency that is not merely the projection of invariance-based reasoning onto a collinear dataset. No such evidence has been provided. The present paper's contribution, in this context, is to supply what Ganesh leaves deliberately open: a positive account of what $\Omega$ actually is, grounded in autopoietic organisation, metabolic investment, and the identity criterion, and a principled basis for the conclusion that no purely computational system, however mechanistically sophisticated, satisfies it.

This is, in the end, a humanising conclusion as much as a philosophical one. It restores to the question of mind its proper gravity: the gravity of something that is not just calculated, but \emph{lived}.

There is, moreover, a further and practically urgent corollary that follows directly from the foregoing analysis. If human consciousness is not epiphenomenal---if it exerts genuine causal power over the physical world, as the arguments of Popper and Eccles, Penrose and Hameroff, and Froese collectively suggest is at least possible---then no system whose behaviour is exhausted by computation can, even in principle, replicate the full causal profile of a conscious human agent. Computation, as the abstraction fallacy makes clear, is an observer-dependent description of physical processes, not an intrinsic physical power in its own right; it cannot, therefore, be the vehicle through which consciousness acts on the world. The consequences for the project of artificial general intelligence are correspondingly stark. Systems such as large language models may exhibit striking and, in many domains, genuinely impressive aspects of human-like intelligence: they can reason, compose, translate, summarise, and produce outputs that are, in a wide range of practical contexts, indistinguishable from those of competent human agents. Recent work by Griffiths et al.\ has argued persuasively that modern neural networks now exhibit compositionality, productivity, and rapid learning at levels previously thought to require symbolic processing, and raises the possibility that such systems may implement or closely approximate symbolic cognition at what Marr would call the algorithmic level.~\citep{Marr1982, GriffithsEtAl2026} This is a significant finding about cognitive architecture, and the present paper does not dispute it. What it does dispute is the inference from behavioural or cognitive equivalence to phenomenal equivalence. Establishing that a system approximates symbolic cognition at the algorithmic level is not the same as establishing that it instantiates consciousness, for the reasons canvassed throughout this paper: computational equivalence, whether symbolic or subsymbolic, does not entail physical identity, and it is physical identity---not representational fidelity---that the instantiation of consciousness requires. But there remain dimensions of human capability---those grounded in embodied, metabolically invested, autopoietically organised existence, and in the genuine causal efficacy that consciousness, on the present account, may contribute to action---that are not merely practically difficult but \emph{in principle} inaccessible to systems whose operation reduces to formal symbol manipulation.\footnote{This in-principle inaccessibility is not merely a philosophical claim: it has mathematical grounding in Rosen's category-theoretic proof that anticipatory systems are irreducible to mechanisms. Even if every component of a living system admits a complete mechanistic-computational description, the anticipatory system as a whole cannot be recovered from those descriptions, because the relevant properties belong to the relational organisation of the whole and are destroyed by decomposition. A formal symbol manipulator, however sophisticated, is a mechanism in Rosen's sense: its organisation is imposed from without and fully capturable by efficient causation. No assembly of such mechanisms can produce the closure to efficient causation that constitutes a living, anticipatory, conscious system. The impossibility is structural, not contingent.} AGI, understood as the full replication of human intelligence across all domains of human life and activity, cannot be realised by computation alone. This is not a counsel of despair about artificial intelligence, nor a denial of the remarkable and expanding range of tasks that computational systems can perform with distinction. It is, rather, a principled demarcation: one that takes seriously both the achievements of artificial intelligence and the irreducible depth of what it means to be a minded, embodied, mortal creature for whom the world genuinely matters.

\end{document}